\documentclass[preprint,12pt,numbers,sort&compress]{elsarticle}

\usepackage{amssymb}
\usepackage{setspace}
\usepackage{graphics}
\usepackage{graphicx}
\usepackage{amsfonts}
\usepackage{amssymb}
\usepackage{amsmath}
\usepackage{graphicx}
\usepackage{bm}
\usepackage{color}
\usepackage{epsfig}
\usepackage{calc}
\usepackage{ifthen}
\usepackage{subfigure}
\usepackage{graphicx}
\usepackage{epstopdf}
\usepackage{epsfig}
\journal{Carbon}

\begin{document}
\doublespacing
\begin{frontmatter}



%\title{\textbf{Over-barrier side-band electron emission from graphene with time-oscillating potential}}

\title{Adiabatic control of surface plasmon-polaritons in a 3-layers graphene curved configuration}
%\tnotetext[label1]{Fax: +65 6499 4558}
\author{Wei Huang$^{\ast 1}$}
 %\ead{wei_huang@mymail.sutd.edu.sg}
\author{Shi-Jun Liang$^{\ast 1,2}$}
\author{Elica Kyoseva$^{3}$}
\author{L. K. Ang\corref{cor2}\fnref{label1}}
 \ead{ricky\_ang@sutd.edu.sg}
\cortext[cor1]{Equal contribution to this work}
\fntext[label2]{Tel.: +65 6499 4558}
\cortext[cor2]{Corresponding author}
\address{$^1$Engineering Product Development, Singapore University of Technology and Design, Singapore 138682.\fnref{label1}}
\address{$^2$National laboratory of solid state microstructures, School of physics, Collaborative Innovation Center of Advanced Microstructures, Nanjing university, Nanjing 210093, China. \fnref{label2}}
\address{$^3$Institute of Solid State Physics, Bulgarian Academy of Sciences, 72 Tsarigradsko Chaussee, 1784 Sofia, Bulgaria}


%\author{Shi-Jun Liang$^\dag$, L. K. Ang$^{\dag,\ddag}$\footnote{Tel.: Fax: +65 6499 4558. \\E-mail: ricky\_ang@sutd.edu.sg(L. K. Ang)}, S. Sun$^\ddag$}

%\address{$^\dag$Engineering Product Development, Singapore University of Technology and Design, Singapore 138682.}
%\address{$^\ddag$School of Electrical and Electronic Engineering, Nanyang Technological University, Singapore 639798.}

\begin{abstract}
In this paper, we utilize coupled mode theory (CMT) to model the coupling between surface plasmon-polaritons (SPPs) between multiple graphene sheets.
By using the Stimulated Raman Adiabatic Passage (STIRAP) Quantum Control Technique, we propose a novel directional coupler based on SPPs evolution in three layers of graphene sheets in some curved configuration.
Our calculated results show that the SPPs can be transferred efficiently from the input graphene sheet to the output graphene sheet, and the coupling is also robust that it is not sensitive to the length of the device configuration's parameters and excited SPP’s wavelength.
\end{abstract}

\end{frontmatter}

% \linenumbers


\section{Introduction}
The optical properties of graphene have attracted a large number of attentions from researchers since its first mechanical exfoliation in 2004, in particular for Surface Plasmon-Polaritons (SPPs), which are evanescent electromagnetic waves coupled with free electron plasma oscillations \cite{Gong15}. SPPs can propagate on the surface of metal or graphene \cite{Sanderson15, Luo13, Li17}.
The wavelength of SPPs on metal is in the visible spectrum, while the wavelength of SPPs on graphene is in the near-infrared spectrum \cite{Li14}.
There are some unique advantages of SPPs supported by graphene comparing with those on metals.
Firstly, the SPPs on graphene are more highly confined on the graphene surface than metals \cite{Barnes04, Williams08, Zayats05, Ditlbacher02, Ooi16}.
Due to the low and saturable absorption as well as weak electron-phonon interaction, the damping loss of SPPs supported by graphene is relatively low \cite{Tassin12, Koppens11}, and the propagation length of SPPs on graphene is relatively longer than SPPs propagation on metal's surface.
Typically, the propagation length of SPPs on graphene could reach dozens of wavelength in the infrared and terahertz region \cite{Wang12,Jablan09, Luo13}.
For graphene, its tunable Fermi level via gating also enables the dynamical control of propagation properties of SPPs.
Thus, graphene-based SPPs may find much wider applications in integrated optics and other fields associated with optics \cite{Bludov12,Salihoglu12}.

The SPPs on graphene can be treated as analogous to optical waveguide, due to SPPs propagation guiding by graphene surface.
SPPs can be coupled together between two parallel plane graphene sheets \cite{Wang12,Wang122,Jiang16}.
One advantage of SPPs on graphene as compared to optical waveguide is that the associated devices could be extremely smaller in length scale than optical waveguide.
With this feature, SPPs on graphene could be remarkable useful in integrate optics and nanophotonics, such as applications in optical communication, optical computation and optical quantum computation.
However, the setups of devices proposed in previous works \cite{Wang12, Christensen11, Abajo14} are based on the coupling of plasmons of two separated graphene layers. The coupling is very sensitive to the device's geometric structure (e.g. device length, distance between graphene layers) and excited SPP's wavelength. As a consequence, the performance of proposed devices is not robust against variation in device's geometric structure and excited SPP's wavelength, which is undesirable for practical applications.
In this paper, we introduce the Stimulated Raman Adiabatic Passage (STIRAP) Quantum Control Technique \cite{Vitanov01, Vitanov17, Huang17} into graphene SPPs coupling, with improving the propagation of SPPs to be more robust against varying geometrical parameters.
STIRAP is a robust three-level coherent quantum control with designing two Gaussian shape coupling strengths, which can realize complete transfer from initial state to final state without any population in middle state.
The advantage of utilizing STIRAP technique in designing the coupling device of graphene SPPs is that one can realize complete transfer of SPPs energy from top to bottom layer with robust against all device's geometric structure (e.g. device length, curvature radius $R$, $\delta$) and excited SPP's wavelength.
In the present paper, we design a novel robust device based on SPPs coupling among three graphene sheets with a unique curved configuration (see Fig. 2).
Our calculations show that our device is able to be completely and robust transferred from input graphene sheet to output graphene sheet with varying wavelength of SPPs and geometry layout parameters (see Fig. 4).

\section{Model}
In our model, we employ the coupled mode theory (CMT) to describe SPPs coupling between graphene sheets.
CMT is widely used theory in describing coupling between two optical waveguides, due to the overlap of their evanescent electromagnetic fields. 
The SPP propagates like photon in the conventional optical waveguide. The evanescent wave of SPPs on each graphene layer can overlap each other. Thus the energy transfer can be realized between two graphene layers via optical tunneling. And the coupling between two graphene sheets can be described by CMT. Recently, there have two remarkable papers to describe coupling of SPPs on separated graphene layers by using CMT \cite{Wang12,Wang122}.

We first consider only one layer graphene sheet located at $z=0$. The TM polarized SPPs modes are excited on graphene. The electromagnetic fields can be described by \cite{Bludov13} $E = (E_{m,x}, 0, E_{m,z}) e^{iqx} e^{-k_m |z|}$ and $B = (0, B_{m,y}, 0) e^{iqx} e^{-k_m |z|}$, where $m=\{1,2\}$ is the index of the two dielectric mediums on each side of graphene and $q$ is the propagation constant of SPPs.
The evanescent electromagnetic fields of each side of graphene are exponentially decaying with constant $k_{m}$, where the $k_{m}= \sqrt{q^2-\omega^2 \epsilon_{m}/c^2 }$. The $\epsilon_{m}$ is the permittivity of medium with index $m$ and $\omega$ is the frequency of incident light in air.

We obtain the dispersion relation \cite{Bludov13}, by substituting the electromagnetic fields on graphene into the Maxwell equations with appropriate boundary conditions, which is given by $\frac{\epsilon_1}{k_1} + \frac{\epsilon_2}{k_2} +i \frac{\sigma_g}{\epsilon_0 \omega} =0$. The $\sigma_g$ is the surface conductivity of graphene, which is given by the Drude formula $\sigma_g = \sigma_0 \frac{4 E_{F}}{\pi} \frac{1}{\hbar \gamma - i\hbar \omega}$ without considering interband contribution.
Here, we have $\sigma_0 = \pi e^2 /(2 h)$, the relaxation rate, $\gamma=2\pi (ev_{F}^2)/(\mu_{e} E_{F})$ \cite{Ooi16, Ooi17}, where $v_F=10^6$ m/s is the Fermi velocity, $\mu_e=6 \times 10^4 cm^{2}/Vs$ is the graphene's carrier mobility and $E_{F}$ is the Fermi level of graphene.

We numerically solve the dispersion relation to determine the propagation constant $q$ and the frequency of incident light in free space $\omega$, so to calculate the mode of the SPPs excitation on graphene.
Note that SPPs mode of graphene is continuous, unlike optical waveguides, due to the non-reflection property of SPP on the graphene.
The profile of the SPPs mode $u_{m}(z)$ is given by $u_{m}(z)=E_{m,z}exp(-k_{m}|z|)$, where $k_{m}$ is the calculated exponentially decaying constant.
For simplicity, following Eq. (1) to Eq. (3) have been normalized, which a normalized parameter is $N=\sqrt{\int^{+\infty}_{-\infty} |u_m(z)|^2 dz}$.

In our model, we make the notation $\Psi_1(x,z)$ ($\Psi_2(x,z)$) as the wave function of SPPs on one (another) graphene sheet, written as
\begin{equation}
\begin{aligned}
\Psi_1(x,z)= a_{1}(x) u_{1}(z) \exp(-i q x), \\
\Psi_2(x,z)= a_{2}(x) u_{2}(z) \exp(-i q x),
\end{aligned}
\end{equation}
where $a_{1}(x)$ and $a_{2}(x)$ are the amplitudes of the modes with respect to SPPs on two graphene sheets. The mode profiles of SPPs on two graphene sheets $u_{1}(z)$ and $u_{2}(z)$ are determined by the dispersion equation. We take the notation as $\psi_{1}$=$u_{1}(z) \exp(-i q x)$ and $\psi_{2}$=$u_{2}(z) \exp(-i q x)$, where $\psi_{1}$ and $\psi_{2}$ must be satisfied by Helmholtz equations in the $x$ direction.
%\begin{equation}
%\begin{aligned}
%\dfrac{\partial^2}{\partial x^2} \psi_{1} + q^2 \psi_1=0, \\
%\dfrac{\partial^2}{\partial x^2} \psi_{2} + q^2 \psi_2=0.
%\end{aligned}
%\end{equation}
Based on the CMT model, we can manipulate the Helmholtz equations with the source terms to obtain
\begin{equation}
\begin{aligned}
\dfrac{\partial^2}{\partial x^2} \Psi_{1}(x,z) + q^2 \Psi_{1}(x,z) = -(k_{2}^2 - k_0^2)\Psi_{2}(x,z), \\
\dfrac{\partial^2}{\partial x^2} \Psi_{2}(x,z) + q^2 \Psi_{2}(x,z) = -(k_{1}^2 - k_0^2)\Psi_{1}(x,z),
\end{aligned}
\end{equation}
where $k_0=\sqrt{q^2 - \omega^2 \epsilon_{g} / c^2}$ with effective graphene permittivity $\epsilon_{g}=1+i \sigma_g \eta_0 c / (w \Delta)$. Here,  $\sigma_g$ is the surface conductivity of graphene by Drude formula, $\eta_0 (\approx 377 \Omega)$ is the free space impedance and $\Delta$ (typically 0.33nm) is the thickness of the single layer graphene \cite{Merano16}. These equations are consistent with the conventional optical waveguide coupled equations \cite{Saleh91}.

By substituting the wave functions of the SPPs on two graphene sheets into the given Helmholtz equations, we simplify the formation by using the slowly varying envelope approximation \cite{Saleh91}, namely $\frac{d^2 a_1}{dx^2} \ll \frac{d a_1}{dx}$ and $\frac{d^2 a_2}{dx^2} \ll \frac{d a_2}{dx}$.
Under this approximation, the coupling equations can be rewritten as a Schr\"odinger-like equation of a two-level system, given by
\begin{equation}
i\dfrac{d}{d x}
\begin{bmatrix}
a_{1} \\
a_{2}
\end{bmatrix}
= \begin{bmatrix}
0 & C_{12} \\
C_{21}  & 0
\end{bmatrix} \begin{bmatrix}
a_{1} \\
a_{2}
\end{bmatrix}.
\end{equation}
Here, $C_{12}$ and $C_{21}$ are the coupling coefficients: $C_{12} = \frac{1}{2} \frac{k_{2}^2 - k_0^2}{q}  \int^{+\infty}_{-\infty} u_{1}(z) u_{2}(z) dx$ and $C_{21} = \frac{1}{2} \frac{k_{1}^2 - k_0^2}{q}  \int^{+\infty}_{-\infty} u_{1}(z) u_{2}(z) dx$.

In our first example as shown in Fig. 1 (a), we consider that the wavelength of the incident light is 10$\mu m$.
Two parallel graphene sheets are placed at $z=d/2$ and $z=-d/2$ respectively, where $d$ is the spacing between the two graphene sheets.
The substrate and its surrounding material is assumed to be $SiO_2$, with the dielectric constant of $\epsilon_h = 3.9$.

\begin{figure}[hbtp]
\centering
\includegraphics[width=1\textwidth]{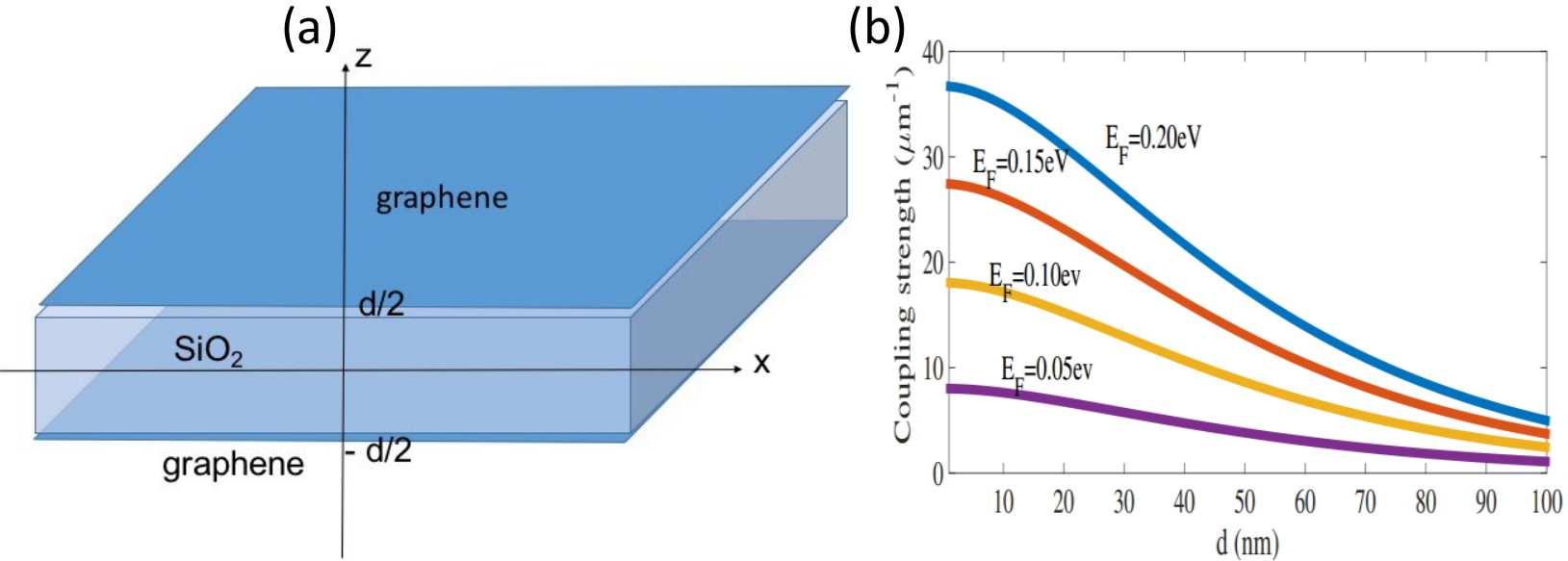}
\caption{(a)The scheme of two parallel graphene sheets with surrounding material $SiO_2$, for which dielectric constant is 3.9. $d$ is distance between two graphene sheets. (b)The coupling coefficient of two parallel graphene sheets with respect to distance $d$ and with different Fermi level $E_{F}$.}
\end{figure}

Due to symmetry in this example, the coupling coefficients are equal: $C_1=C_2$.
According to our model, the coupling strengths are shown in Fig. 1 (b) as a function of distance $d$ up to 100 nm for different Fermi level = $E_F$ = 0.05, 0.1, 0.15 and 0.2 eV.
From the results, we can see that the coupling coefficient reduces significantly with increasing $d$, which is consistent with the previous work \cite{Wang12}.
However, the magnitude in coupling strength is larger than previous work for the same values of $d$, because of the different process in exciting and coupling SPPs.
Note we only consider the excitation of SPPs on the input graphene sheet, instead of both graphene sheets. Thus, we have larger intensity of SPPs, which leads to larger coupling strength.
For example, at $d=20 nm$, our results show coupling strength of 24 $\mu m^{-1}$, which is larger than the reported 19 $\mu m^{-1}$.

In the first example, we only consider the coupling between two separated graphene sheets.
In some practical applications, however, the coupling among multiple channels may become prominent, specially for optical devices analogous to wavelength division multiplexing techniques.
To account for multiple layers $n$ larger than 2, Eq. (3) can be easily extended to determine the coupling among $n$ ($>$2) layers of graphene sheets, which becomes

\begin{equation}
i\dfrac{d}{d x}
\begin{bmatrix}
a_{1} \\
\vdots \\
a_{n}
\end{bmatrix}
= \begin{bmatrix}
0 & \Omega_1(x) & \ddots \\
\Omega_1(x) & \ddots & \Omega_{n-1}(x) \\
\ddots & \Omega_{n-1}(x) & 0 \\
\end{bmatrix} \begin{bmatrix}
a_{1} \\
\vdots \\
a_{n}
\end{bmatrix}.
\end{equation}
Here, $\Omega_1(x)$ is the coupling SPPs between first and second layer graphene sheets and $\Omega_{n-1}(x)$ is the coupling SPPs between $(n-1)^{th}$ and $n^{th}$ layer graphene sheets. Similarly, $a_1$ and $a_n$ are the SPPs amplitudes of first and $n^{th}$ graphene layer.

\begin{figure}[hbtp]
\centering
\includegraphics[width=1\textwidth]{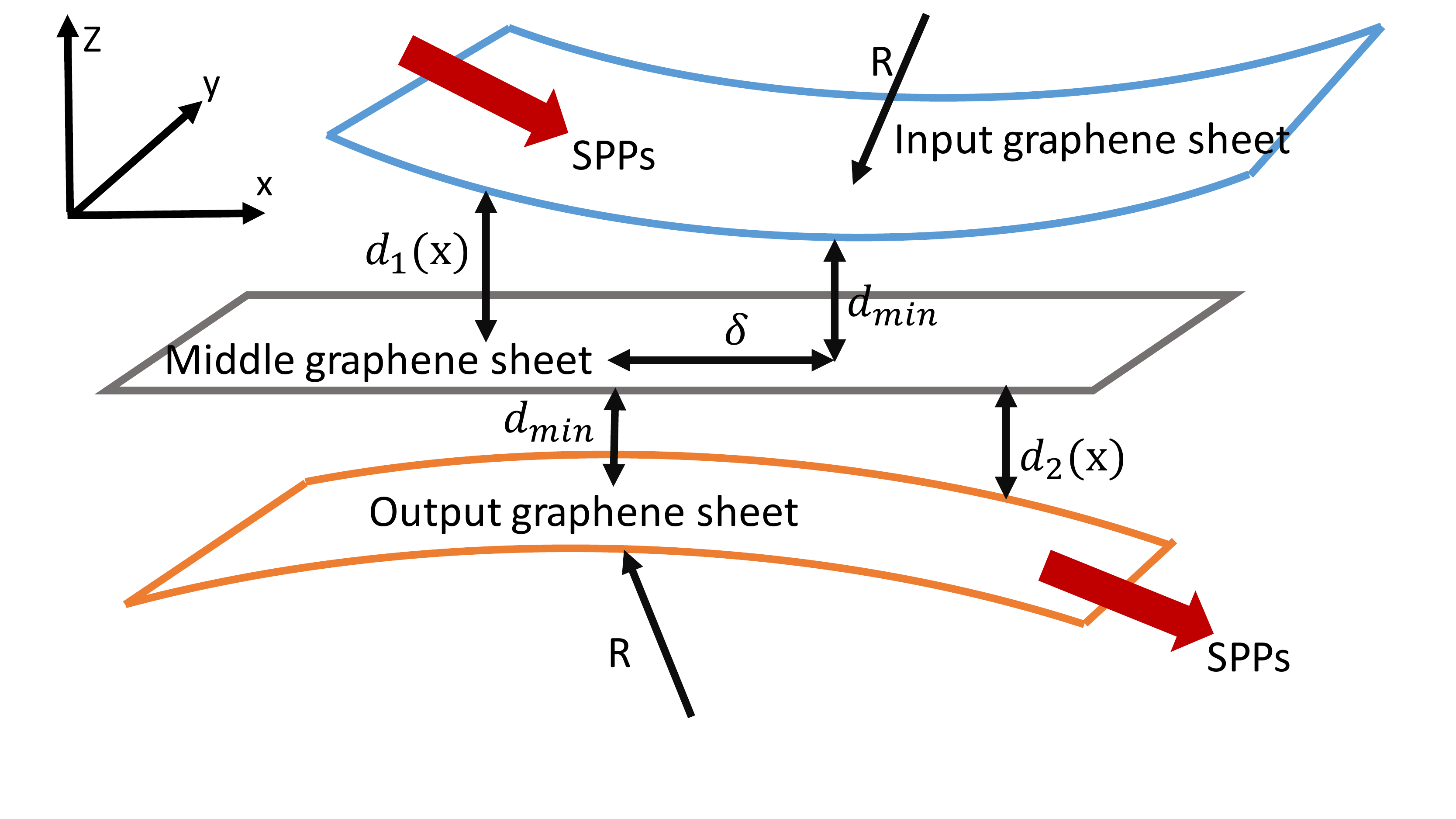}
\caption{The scheme of adiabatic device of SPPs transferring from input to output graphene sheet. $R$, $\delta$ and $d_{min}$ are the geometry parameters. The middle graphene sheet is placed at $z=0$ plane. The input and output graphene sheets are weakly curved, with opposite curvature of radius $R$, which are transverse displaced from each other by a value $\delta > 0$. The minimum distance $d_{min}$ is longitudinal shortest distance between input and middle (middle and output) graphene sheets in the $x-z$ plane.}
\end{figure}

In a previous design based on SPPs on graphene \cite{Wang12}, the work only considered two parallel graphene sheets and its operation is extremely sensitive to the coupling length and wavelength of SPPs.
To overcome this limitation, we devise an adiabatic SPPs coupling device by considering three layers of graphene sheets with the curvature configuration for the top and bottom graphene layers, as shown in Fig. 2.
The excited SPPs on the input graphene sheets (top layer) tunnels to the output graphene sheet (bottom layer) through middle graphene sheet located at $z=0$ plane. Based on calculation, we find that the field of SPP on the each graphene layer can permeate to the surrounding materials within a distance of 23 nm as the field amplitude decays to 1/e of the original intensity at the graphene based on parameters used in our work. Therefore the interlayer space between top (or bottom) and middle graphene layers is close enough to allow optical tunneling in the current configuration. This argument can also be supported through previous literature \cite{Wang12}.
The input and output graphene sheets are weakly curved, with opposite curvature of radius $R$, and they can be transversely displaced from each other by a distance of $\delta > 0$.
The minimum spacing $d_{min}$ refers to the longitudinal shortest distance between the curve graphene sheet to the middle sheet.
The spatial dependence of the spacing $d_1(x)$ and $d_2(x)$ of the input and output graphene sheets with respective to the middle layer are given by $d_1(x)=\sqrt{R^2-(x-\delta/2)^2}+(d_{min}+R)$ and $d_2(x)=\sqrt{R^2-(x+\delta/2)^2}-(d_{min}+R)$.
Note it has been shown that the propagation of SPPs on a slightly bend graphene is nearly same to a perfectly planar graphene sheet \cite{Xiao15}.

Utilizing Fig. 2 as an example, the coupling among three graphene sheets can be determined by Eq. (3) with $n$ = 3, which is similar to Schr\"odinger-like equation of a there-level system:
\begin{equation}
i\dfrac{d}{d x}
\begin{bmatrix}
a_{1} \\
a_{2} \\
a_{3}
\end{bmatrix}
= \begin{bmatrix}
0 & \Omega_1(x) & 0 \\
\Omega_1(x) & 0 & \Omega_2(x) \\
0 & \Omega_2(x) & 0 \\
\end{bmatrix} \begin{bmatrix}
a_{1} \\
a_{2} \\
a_{3}
\end{bmatrix},
\end{equation}
where $\Omega_1$ ($\Omega_2$) represents coupling strength between input and middle graphene sheet (between middle and output graphene sheet) with respect to $x$, such that $\Omega_1(x)$=$C_{12}(x)$=$C_{21}(x)$ and $\Omega_2(x)$=$C_{23}(x)$=$C_{32}(x)$.

To realize robust transfer of the SPPs from the input to the output graphene sheet, we introduce the adiabatic quantum control following (STIRAP) in three-levels-like quantum system.
According to STIRAP theory, we can turn a normal states to an adiabatic states, which are the superposition of normal states.
By appropriate control of associated parameters labelled in Fig. 2, we are able to make $\Omega_1(x)$ and $\Omega_2(x)$ to meet the adiabatic criteria.
In the adiabatic dominated region, fine control over the coupling strengths can enable a complete and robust transferring SPPs from the input graphene sheet to the output graphene sheet.
To achieve the proposed adiabatic control, we can tune the appropriate values of $d_1(x)$ and $d_2(x)$ and to monitor the robustness of the control for a reasonable range of $R$, $\delta$ and $d_{min}$.

In our analysis of Fig. 2, the proposed three layers of graphene sheets are surrounded by $SiO_2$ (dielectric constant is 3.9) and all three layers of graphene sheets are assumed to have the same Fermi level $0.15$ eV.
Incident light of 10 $\mu m$ wavelength is utilized to excite the SPPs on the input graphene sheet.
The corresponding relaxation rate is $\gamma=1.11 \times 10^{12}$ and we choose higher $\gamma=2 \times 10^{12}$ for more practical transport loss of graphene \cite{Wang12,Wang122}, which leads to a maximum propagation length of $L_{x}=4.092$ $\mu m$ by using $L_{x}= 1/ (2\text{Im}(q))$ \cite{Jackson98}.

\section{Results and Discussions}
In the following results, we choose device length of 1 $\mu m$, $R$ = 800 nm, $\delta$ = 200 nm and $d_{min}$ =20 nm as default parameters unless otherwise specified.
Note the device length is chosen as 1 $\mu m$, so it is smaller than the maximum propagation length $L_{x} \approx$ 4 $\mu m$, so that the SPPs can propagate without vanishing all SPPs energy.
While smaller-size device is desirable for integrated optics, but the device length cannot be too small subjected to the adiabatic criteria.

With the two-dimensional nature of graphene, the loss mainly comes from the in-plane scattering, therefore the presence of lossy scattering or damping would not affect the coupling between graphene sheets in the $z$-direction.
In Fig. 3 (a), the coupling strength with loss is almost the same as the coupling strength without loss.
From the calculated coupling strength $\Omega_1(x)$ and $\Omega_2(x)$ for a given geometry setting, we calculate the intensity evolution of SPPs on each graphene layer (input, middle and output), as depicted in Fig. 3 (b).

We make comparison among each intensity evolution of SPPs with and without loss. From Fig. 3 (b), we can see that the SPPs can completely transfer from input graphene sheet ($I_{input}=|a_1|^2$) to output graphene sheet ($I_{output}=|a_3|^2$) without any intensity left in middle graphene sheet ($I_{middle}=|a_2|^2$) at the output of device, even for the presence of loss (dot line).
The transfer of the SPPs excited in the input graphene layer to the output graphene layer through middle one can also be visualized in the evolution process as shown in Fig. 3 (c).
Apparently, there is zero intensity in the middle graphene layer when the transfer is completed.
The length scale for this design is from -300 nm to +300 nm, which is much smaller than previous optical waveguide devices \cite{Paspalakis06, Huang14, Hristova16, Longhi07} and thus possible for more compact photonic devices integration (nanometer photonic device).

\begin{figure}[hbtp]
\centering
\includegraphics[width=1\textwidth]{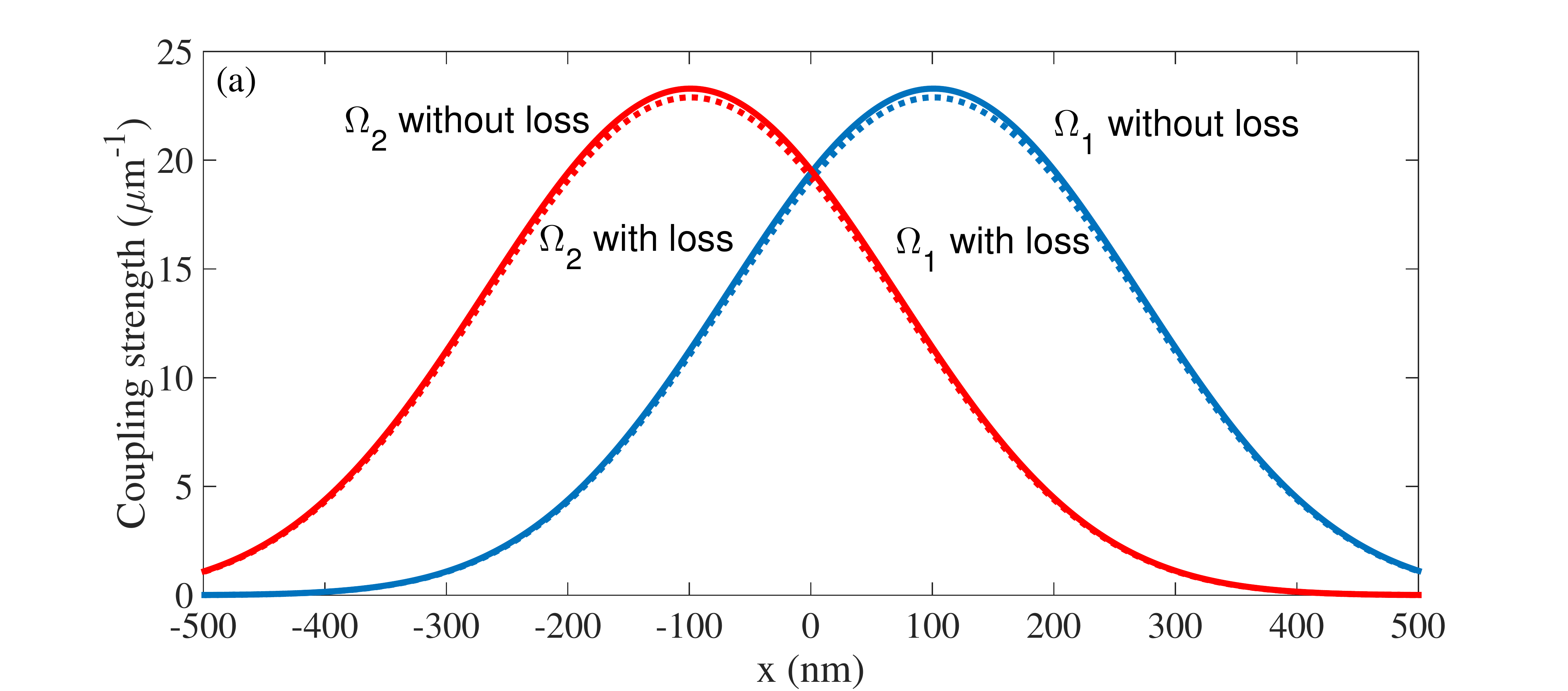}
\includegraphics[width=1\textwidth]{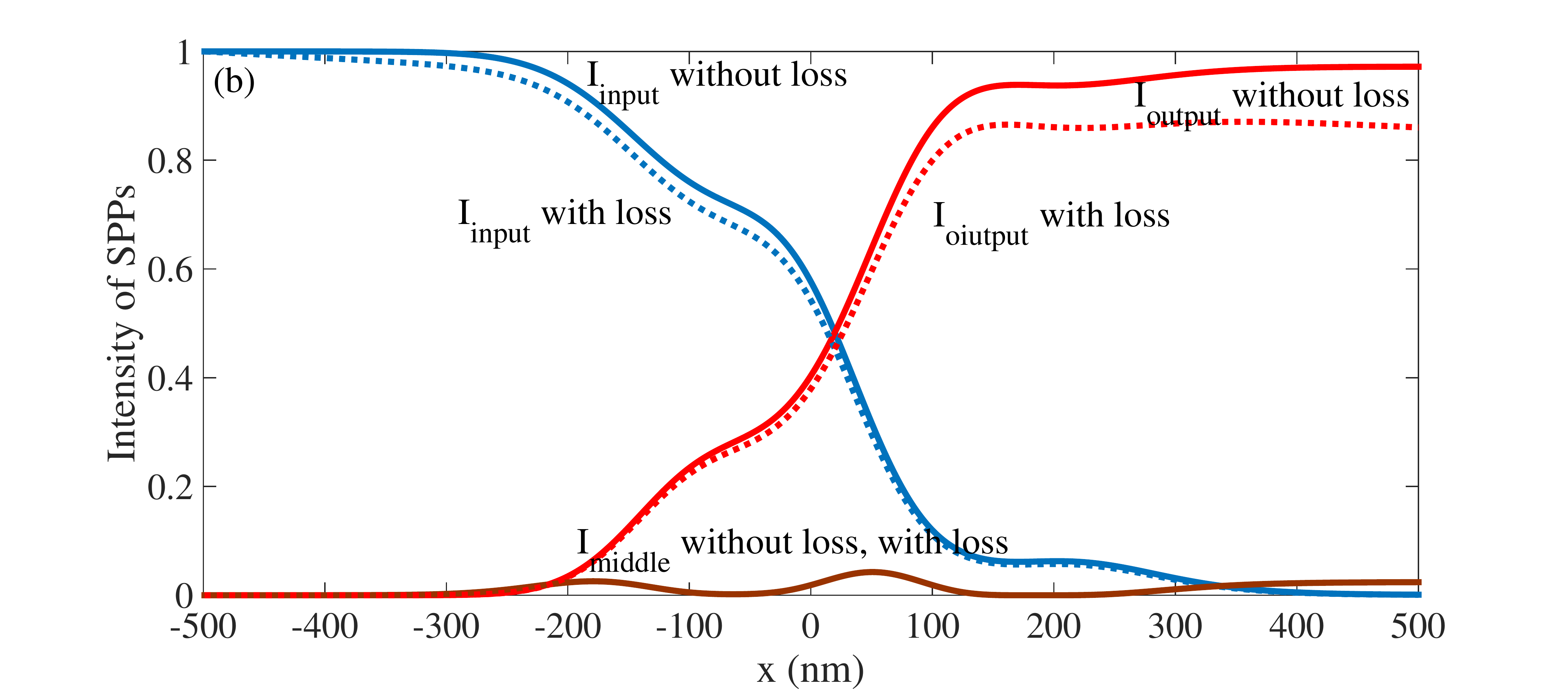}
\includegraphics[width=1\textwidth]{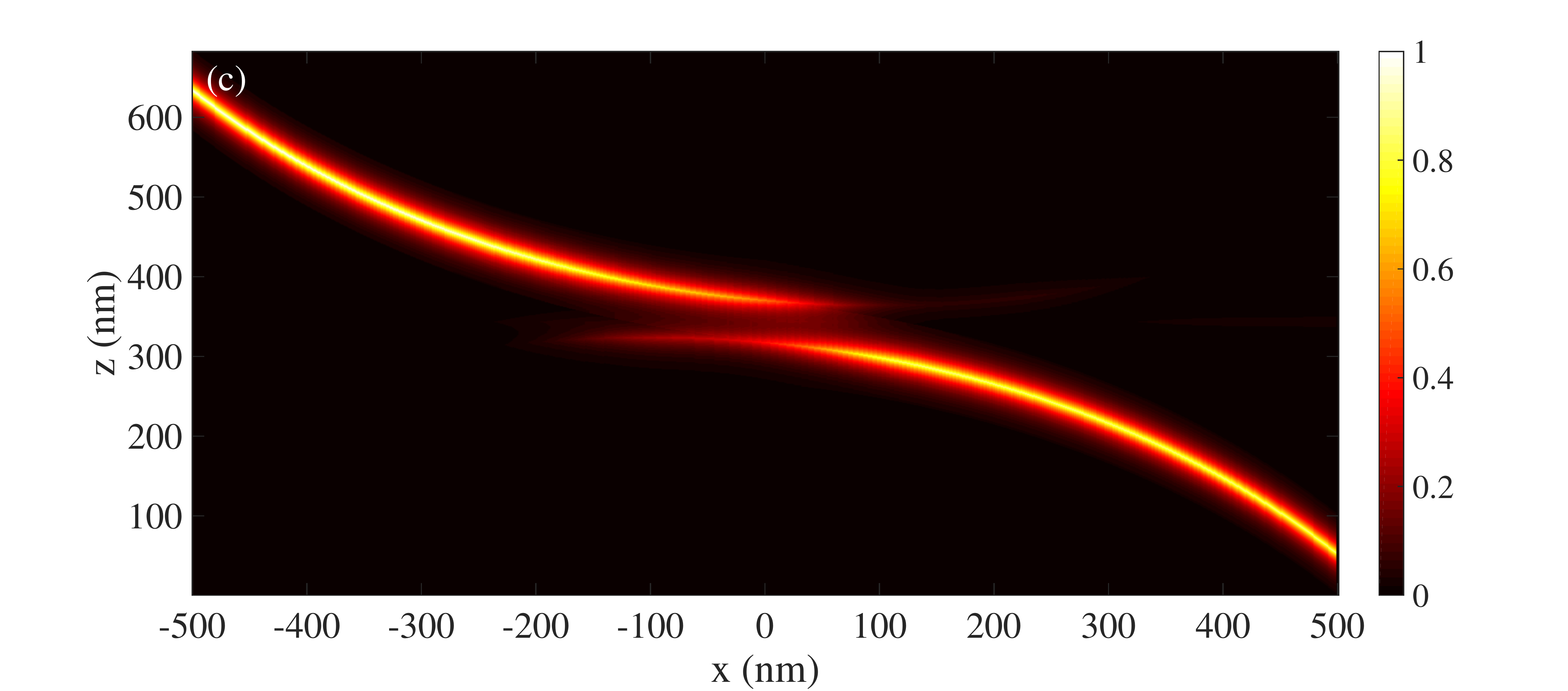}
\caption{(a) The coupling strength $\Omega_1$ ($\Omega_2$) between input and middle graphene sheet (between middle and output graphene sheet) with respect to propagation index $x$. The solid line is SPPs without loss and dot line is SPPs with loss. (b) The intensity of SPPs on input ($I_{input}$), middle ($I_{middle}$) and output ($I_{output}$) graphene sheet with respect to $x$. The solid line only considers situation without loss and dot line considers the situation with loss. (c)The visualizing simulation of SPPs propagating complete transfer from input graphene sheet to output graphene sheet.}
\label{example}
\end{figure}

\begin{figure}[hbtp]
\centering
\includegraphics[width=1\textwidth]{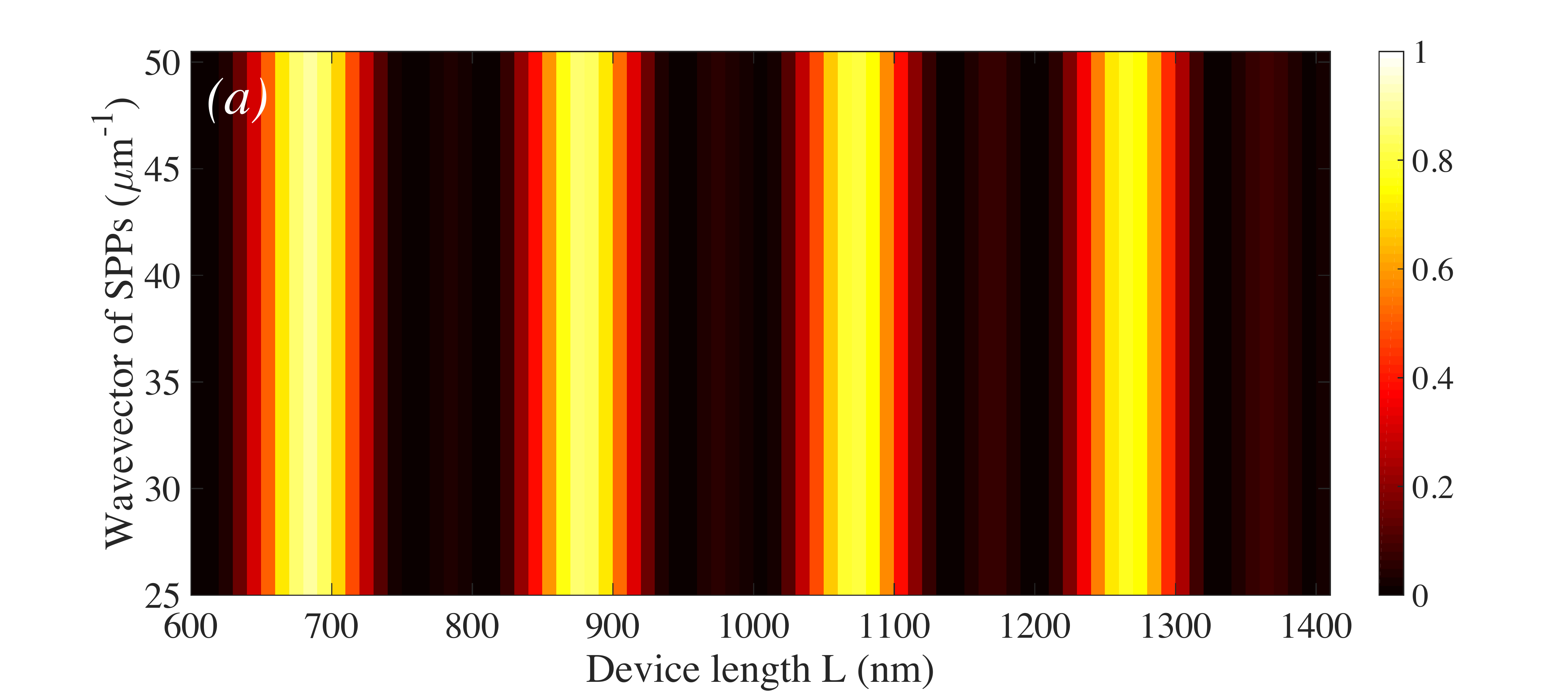}
\includegraphics[width=1\textwidth]{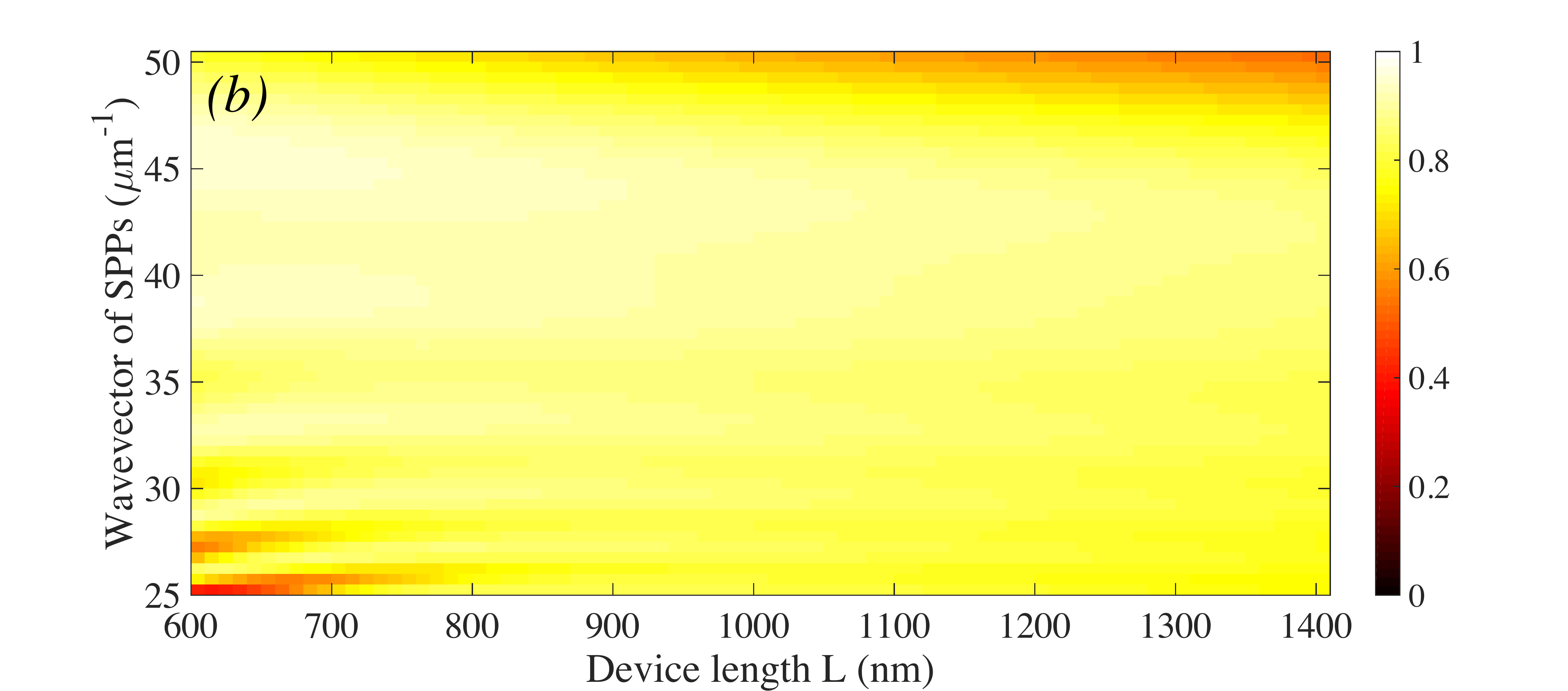}
\includegraphics[width=1\textwidth]{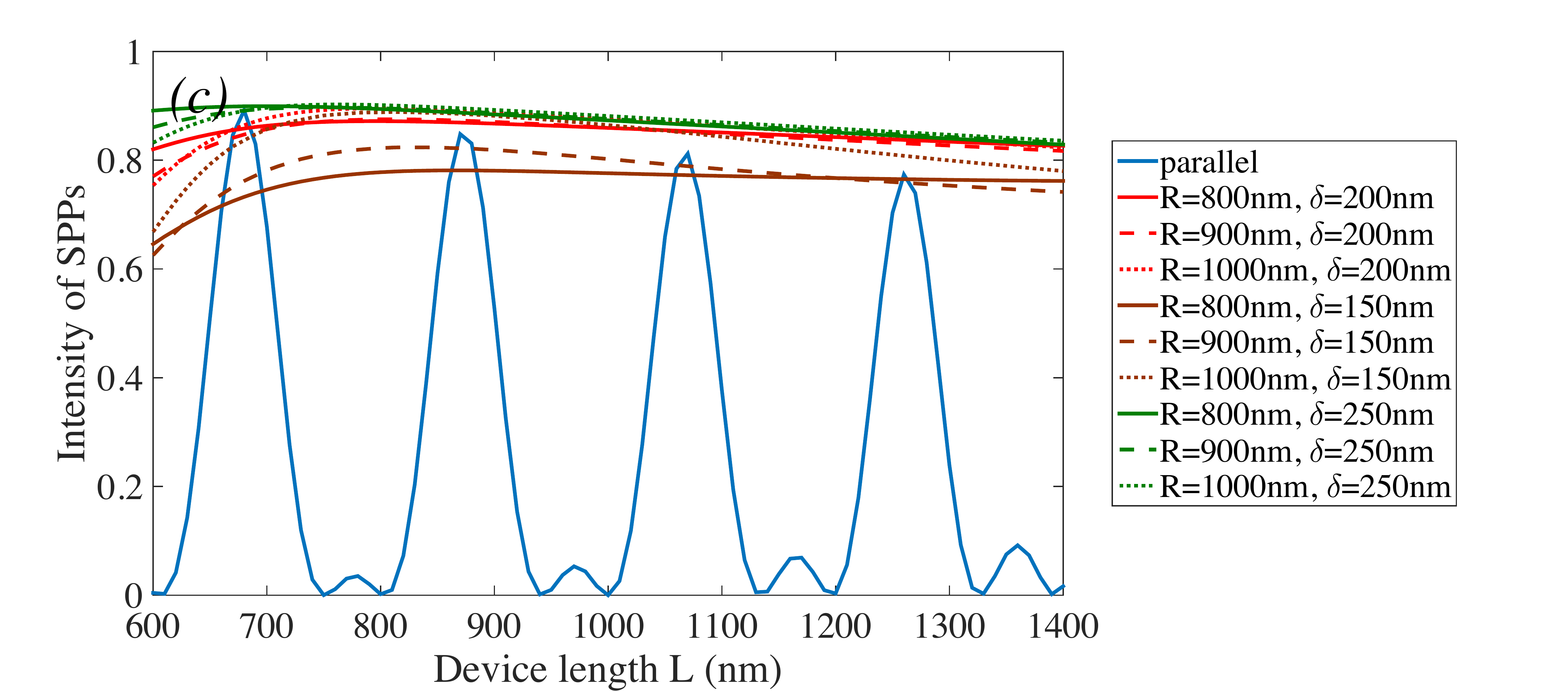}
\caption{Comparison robustness of plane parallel graphene and our device. We set plane parallel graphene with fixed constant $d_{min}=20nm$. For our device, we set fixed the curvature of graphene ($R=800nm$) and $d_{min}=20nm$, $\delta=200nm$. (a) The Intensity of output graphene's SPP with varying wavevector of SPPs and device length $L$ for two SPP parallel graphene device. (b) The Intensity of output graphene's SPP with varying wavevector of SPPs and device length $L$ for our device. (c) We fixed the wavevector as 35$\mu m^{-1}$ (the corresponding wavelength of incident light $\lambda=10 \mu m$) and we compare with parallel graphene device and our device with varying $R$ and $\delta$.}
%\label{Robust}
\end{figure}

As mentioned before, three geometry parameters ($R$, $d_{min}$ and $\delta$) and the SPP excitation frequency ($\omega$) determine the coupling strength functions $\Omega_1(x)$ and $\Omega_2(x)$.
According to STIPAP, it is robust to variation of coupling strength functions $\Omega_1(x)$ and $\Omega_2(x)$, thus our design proposed in this paper should be insensitive to the variations in configuration structure ($R$, $d_{min}$, $\delta$ and $\omega$) and excited SPPs wavelength.
To demonstrate these robustnesses, Fig. 4 (a) first demonstrates the SPPs' intensity on the output graphene sheet of coupling between two parallel graphene sheets (first example), for different wavevector of SPPs and device length $L$.
For 3-layers graphene configuration, Fig. 4 (b) shows the numerical results of the wavevector of SPPs and device length $L$ at fixed R=800 $nm$ and $\delta$=200 $nm$.
Comparing Fig. 4 (a) and Fig. 4 (b), we see that the 3-layers design is more robust with wavevector of SPPs in the range with from 25 $\mu m^{-1}$ to 50 $\mu m^{-1}$ over a wide range of length $L$.
In Fig. 4 (c), we shows the intensity of SPP for a fixed wavevector (35 $\mu m^{-1}$ corresponding to a wavelength of 10 $\mu m$ for different values of $R$, $\delta$ and device length $L$.
From Fig. 4 (c), it is clear than the 3-layer design is more robust (with intensity confined between 0.6 to 0.9) as compared to the 2-layer design with parallel graphene sheet (blue solid line in the figure).

We would like to discuss the experimental realization of our model by showing that the proposed method and range of parameters are within the reach of current technology. Firstly, the synthesis of large area graphene as used in our calculation (device length of 1 $\mu$m) can be achieved though Chemical vapor deposition approach \cite{Xu16}. As often used in the graphene experiment, the Fermi level of $0.15 eV$ can be reached by using the chemical or electrostatic doping \cite{Novoselov04, Zhang05}. Secondly, the curved graphene layer may be considered to employ nanoimprinting process \cite{Xia16} similar to the method described in the literature \cite{Dostalek05}. Finally, we may consider utilize the light to excite the SPP on an extended graphene top layer, by similar to the excitation on the planar graphene configuration. By doing so, the excited SPP can be injected into the top graphene layer.

\section{Conclusion}
In conclusion, we have proposed a new design of graphene SPP based directional coupler by using a 3-layers graphene curved configuration.
By utilizing the coupled mode theory (CMT) and the Stimulated Raman Adiabatic Passage (STIRAP) Quantum Control Technique, we show that the design can serve as a novel and compact adiabatic directional coupler, which is more robust than a typical 2-layer graphen plane configuration.
This finding will be helpful to the future development of more compact and robust integrated optic circuits.

\section{Acknowledgement}
This work is partly supported by Singapore ASTAR AME IRG A1783c0011 and U.S. Air Force Office of Scientific Research (AFOSR) through the Asian Office of Aerospace Research and Development (AOARD) under Grant No. FA2386-17-1-4020. E. K. acknowledges financial support from the European Union’s Horizon 2020 research and innovation programme under the Marie Sklodowska-Curie grant agreement No 705256  - COPQE.

%\bibliographystyle{elsarticle-num}
%\bibliography{reference}

\newpage
\hrule
\section*{Caption}

Caption 1.
(a)The scheme of two parallel graphene sheets with surrounding material $SiO_2$, for which dielectric constant is 3.9. $d$ is distance between two graphene sheets. (b)The coupling coefficient of two parallel graphene sheets with respect to distance $d$ and with different Fermi level $E_{F}$.

Caption 2.
The scheme of adiabatic device of SPPs transferring from input to output graphene sheet. $R$, $\delta$ and $d_{min}$ are the geometry parameters. The middle graphene sheet is placed at $z=0$ plane. The input and output graphene sheets are weakly curved, with opposite curvature of radius $R$, which are transverse displaced from each other by a value $\delta > 0$. The minimum distance $d_{min}$ is longitudinal shortest distance between input and middle (middle and output) graphene sheets in the $x-z$ plane.

Caption 3.
(a) The coupling strength $\Omega_1$ ($\Omega_2$) between input and middle graphene sheet (between middle and output graphene sheet) with respect to propagation index $x$. The solid line is SPPs without loss and dot line is SPPs with loss. (b) The intensity of SPPs on input ($I_{input}$), middle ($I_{middle}$) and output ($I_{output}$) graphene sheet with respect to $x$. The solid line only considers situation without loss and dot line considers the situation with loss. (c)The visualizing simulation of SPPs propagating complete transfer from input graphene sheet to output graphene sheet.

Caption 4. (Color online)
Comparison robustness of plane parallel graphene and our device. We set plane parallel graphene with fixed constant $d_{min}=20nm$. For our device, we set fixed the curvature of graphene ($R=800nm$) and $d_{min}=20nm$, $\delta=200nm$. (a) The Intensity of output graphene's SPP with varying wavevector of SPPs and device length $L$ for two SPP parallel graphene device. (b) The Intensity of output graphene's SPP with varying wavevector of SPPs and device length $L$ for our device. (c) We fixed the wavevector as 35$\mu m^{-1}$ (the corresponding wavelength of incident light $\lambda=10 \mu m$) and we compare with parallel graphene device and our device with varying $R$ and $\delta$.

\newpage
\section*{Figure}
\begin{figure}[h]
\centering
\includegraphics[width=1\textwidth]{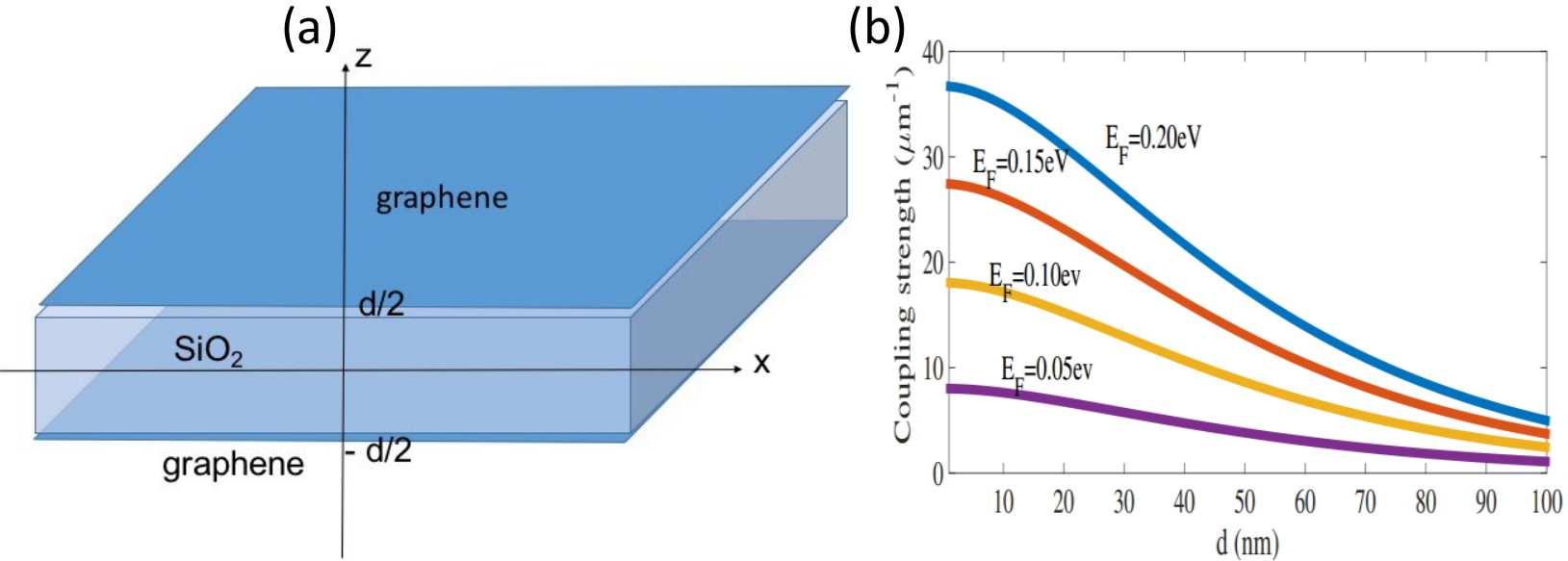}
\begin{center}
Figure 1
\end{center}
%\caption{(a)The scheme of two parallel graphene sheets with surrounding material $SiO_2$, for which dielectric constant is 3.9. $d$ is distance between two graphene sheets. (b)The coupling coefficient of two parallel graphene sheets with respect to distance $d$ and with different Fermi level $E_{F}$.}
\label{device1}
\end{figure}

\begin{figure}[h]
\centering
\includegraphics[width=1\textwidth]{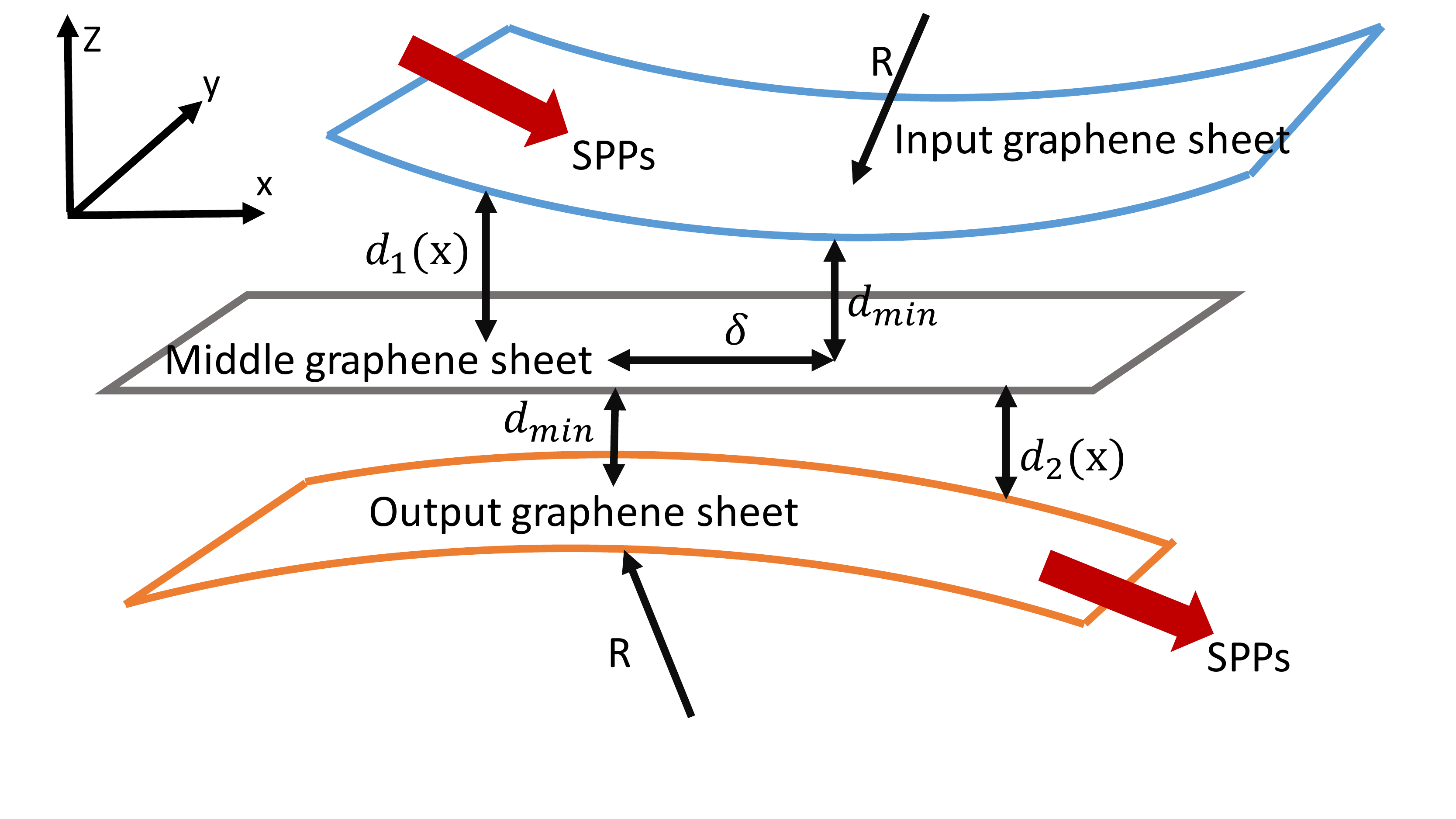}
\begin{center}
Figure 2
\end{center}
%\caption{The scheme of adiabatic device of SPPs transferring from input to output graphene sheet. $R$, $\delta$ and $d_{min}$ are the geometry parameters. The middle graphene sheet is placed at $z=0$ plane. The input and output graphene sheets are weakly curved, with opposite curvature of radius $R$, which are transverse displaced from each other by a value $\delta > 0$. The minimum distance $d_{min}$ is longitudinal shortest distance between input and middle (middle and output) graphene sheets in the $x-z$ plane.}
\label{device2}
\end{figure}

\begin{figure}[h]
\centering
\includegraphics[width=1\textwidth]{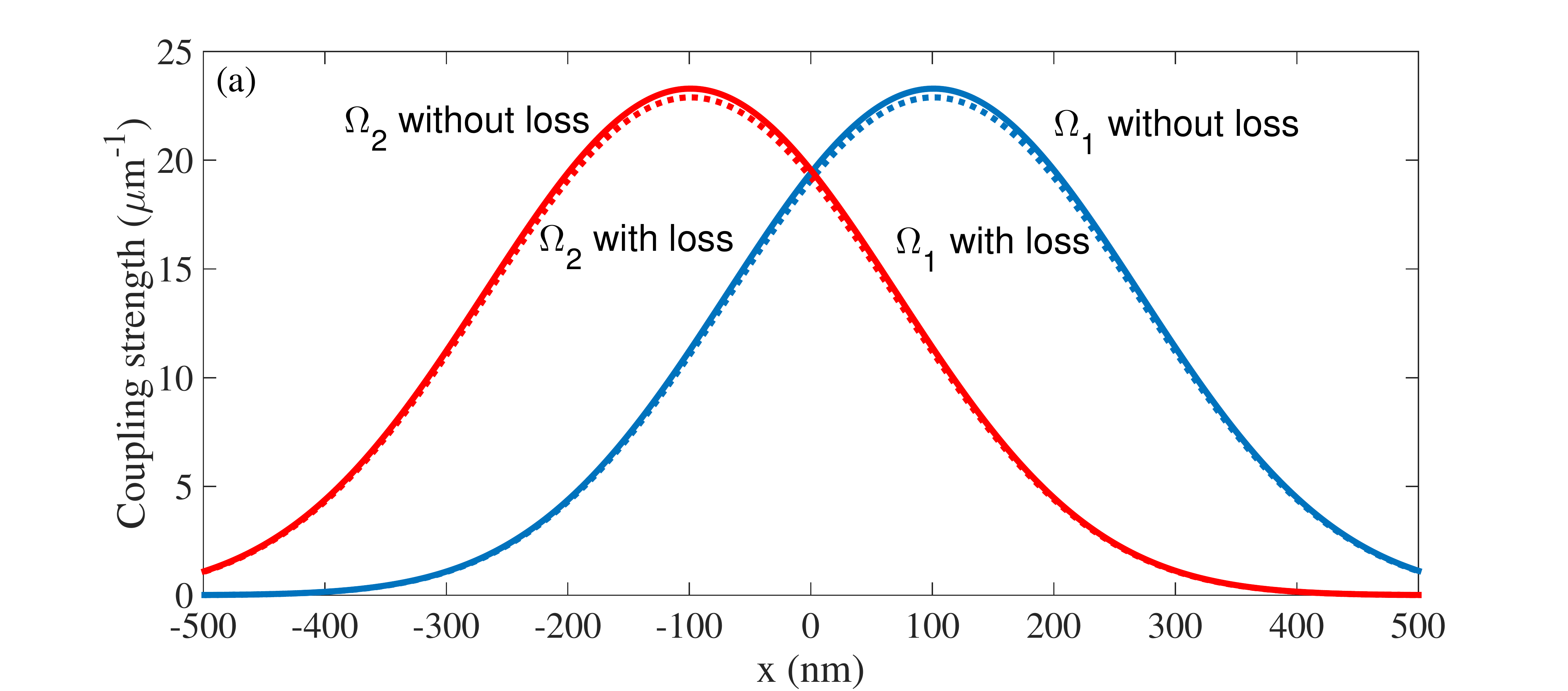}
\includegraphics[width=1\textwidth]{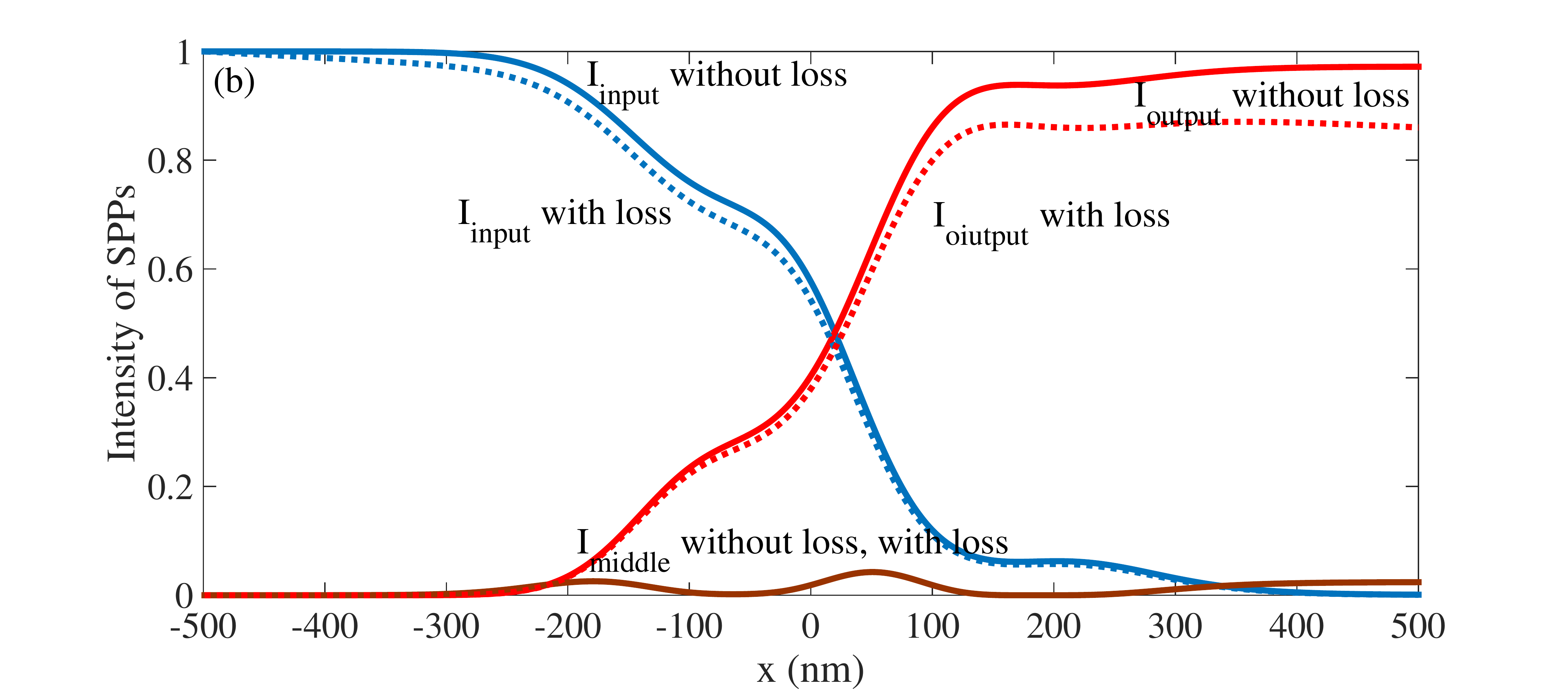}
\includegraphics[width=1\textwidth]{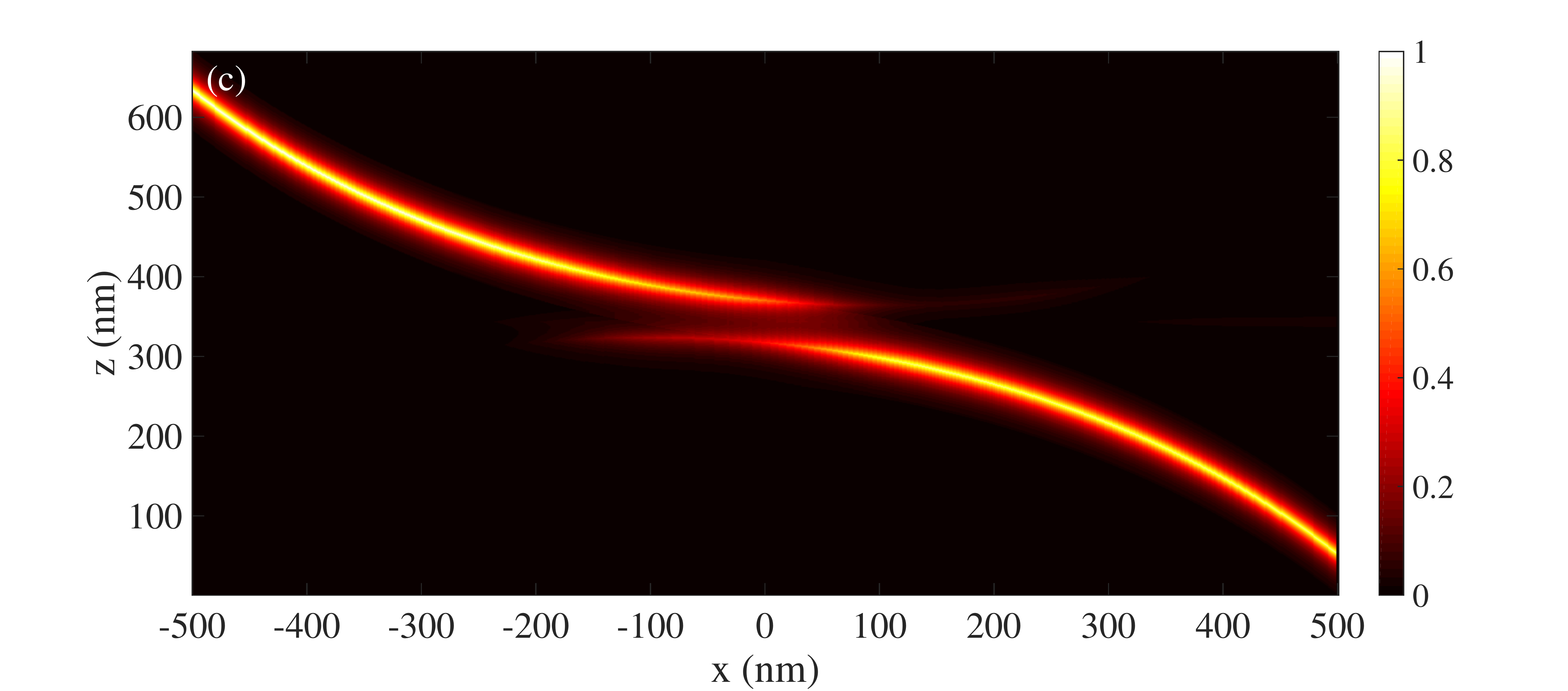}
\begin{center}
Figure 3
\end{center}
%\caption{(a) The coupling strength $\Omega_1$ ($\Omega_2$) between input and middle graphene sheet (between middle and output graphene sheet) with respect to propagation index $x$. The solid line is SPPs without loss and dot line is SPPs with loss. (b) The intensity of SPPs on input ($I_{input}$), middle ($I_{middle}$) and output ($I_{output}$) graphene sheet with respect to $x$. The solid line only considers situation without loss and dot line considers the situation with loss. (c)The visualizing simulation of SPPs propagating complete transfer from input graphene sheet to output graphene sheet.}
\label{example}
\end{figure}

\begin{figure}[h]
\centering
\includegraphics[width=1\textwidth]{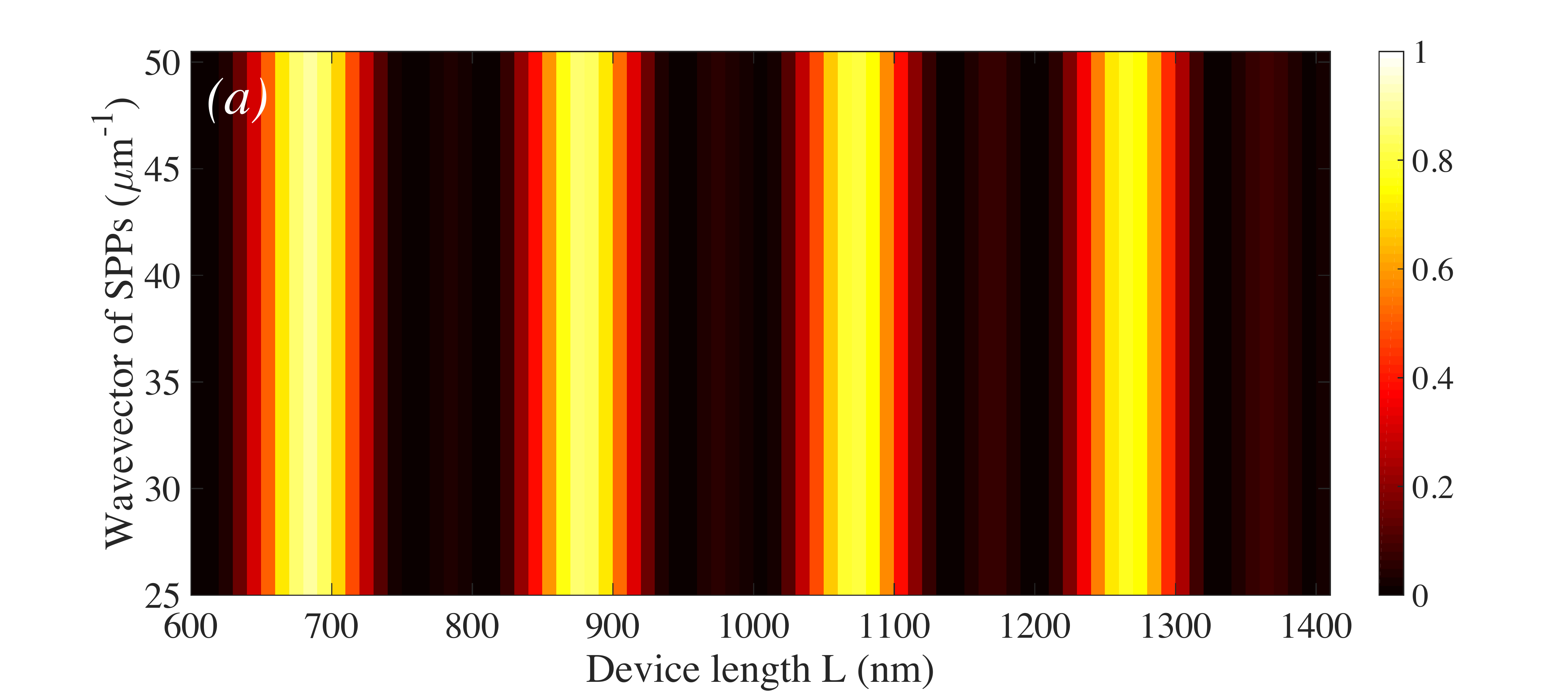}
\includegraphics[width=1\textwidth]{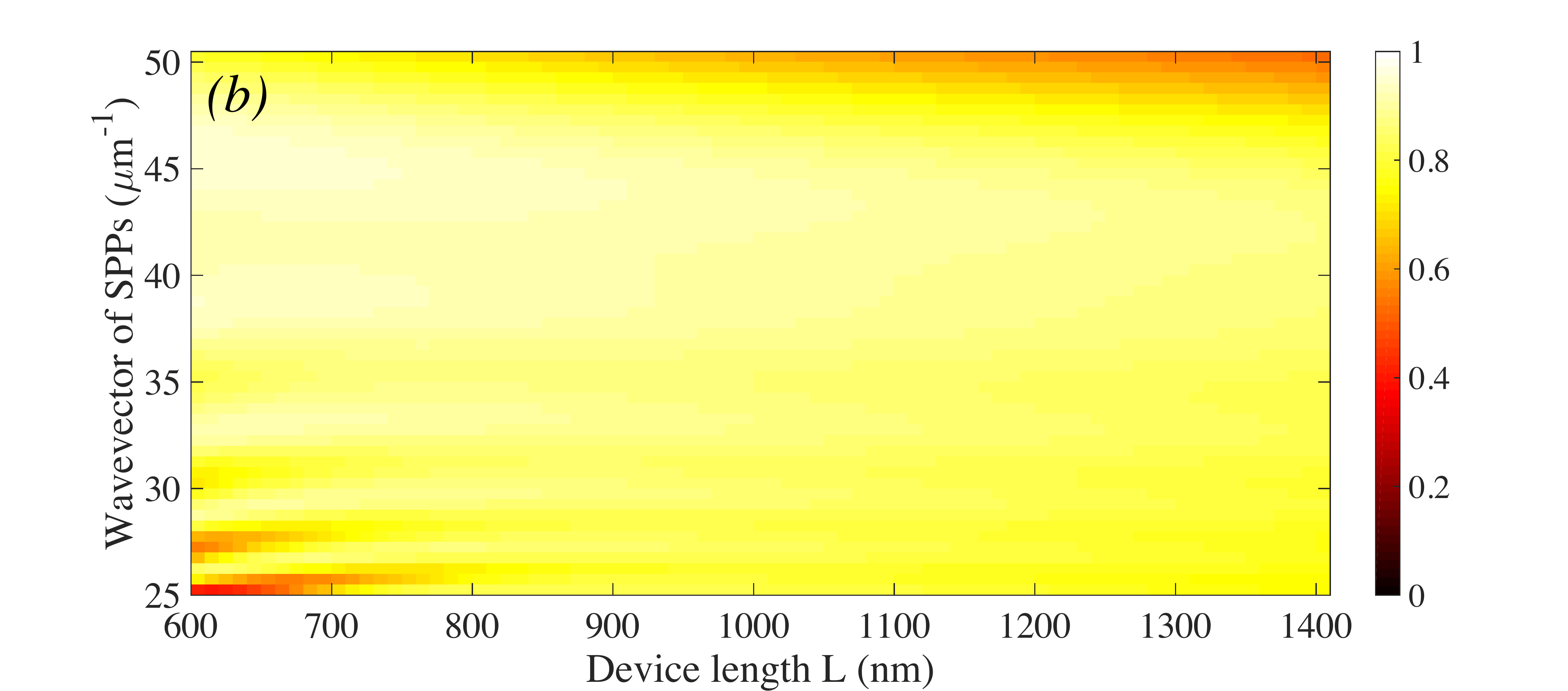}
\includegraphics[width=1\textwidth]{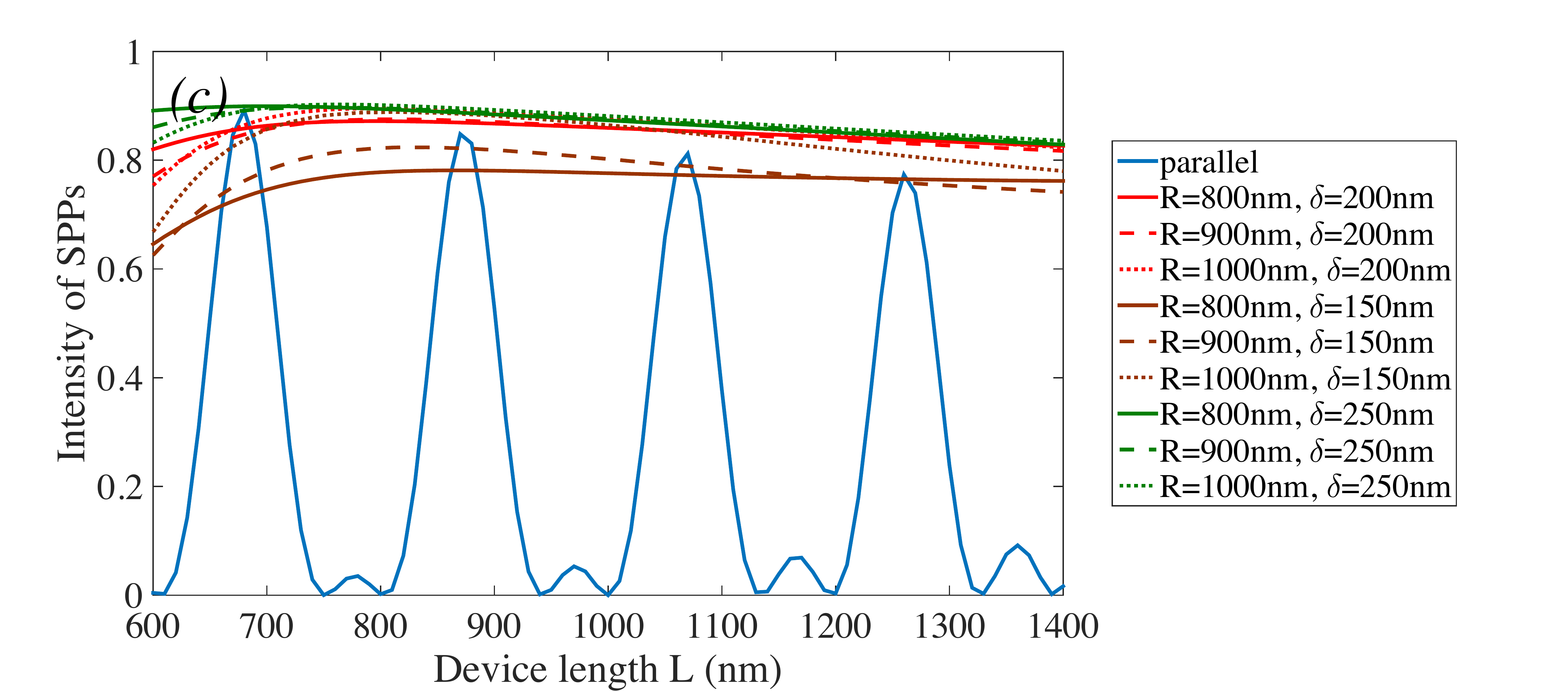}
\begin{center}
Figure 4
\end{center}
%\caption{Comparison robustness of plane parallel graphene and our device. We set plane parallel graphene with fixed constant $d_{min}=20nm$. %For our device, we set fixed the curvature of graphene ($R=800nm$) and $d_{min}=20nm$, $\delta=200nm$. (a) The Intensity of output %graphene's SPP with varying wavevector of SPPs and device length $L$ for two SPP parallel graphene device. (b) The Intensity of output %graphene's SPP with varying wavevector of SPPs and device length $L$ for our device. (c) We fixed the wavevector as 35$\mu m^{-1}$ (the %corresponding wavelength of incident light $\lambda=10 \mu m$) and we compare with parallel graphene device and our device with varying $R%$ and $\delta$.}
\label{Robust}
\end{figure}

\end{document}